\documentclass[12pt]{iopart}

\usepackage{graphicx}
\usepackage{xcolor}
\usepackage{float}
\expandafter\let\csname equation*\endcsname\relax
\expandafter\let\csname endequation*\endcsname\relax
\usepackage{amsmath}
\usepackage{mathtools}
\usepackage{hhline} 
\usepackage{tabularx}
\usepackage{diagbox}
\begin{document}

\title[]{Social network heterogeneity benefits individuals at the expense of groups in the creation of innovation}

\author{F. Zarei$^{1,2}$ , J.  Ryckebusch$^2$, K. Schoors$^1$, \& L. E C Rocha$^{1,2}$}

\address{$^1$ Department of Economics, Ghent University, Ghent, Belgium.}
\address{$^2$ Department of Physics and Astronomy, Ghent University, Ghent, Belgium.}

\ead{Fatemeh.Zarei@UGent.be}
\vspace{10pt}
\begin{indented}

\item[] 
\end{indented}

\begin{abstract}
Innovation is fundamental for development and provides a competitive advantage for societies. It is the process of creating more complex technologies, ideas, or protocols from existing ones. While innovation may be created by single agents (i.e.~individuals or organisations), it is often a result of social interactions between agents exchanging and combining complementary expertise and perspectives. The structure of social networks impacts this knowledge exchange process. To study the role of social network structures on the creation of new technologies, we design an evolutionary mechanistic model combining self-creation and social learning. We find that social heterogeneity allows agents to leverage the benefits of diversity and to develop technologies of higher complexity. Social heterogeneity, however, reduces the group ability to innovate. Not only the social structure but also the openness of agents to collaborate affect innovation. We find that interdisciplinary interactions lead to more complex technologies benefiting the entire group but also increase the inequality in the innovation output. Lower openness to interdisciplinary collaborations may be compensated by a higher ability to collaborate with multiple peers, but low openness also neutralises the intrinsic benefits of network heterogeneity. Our findings indicate that social network heterogeneity has contrasting effects on microscopic (local) and macroscopic (group) levels, suggesting that the emergence of innovation leaders may suppress the overall group performance.
\end{abstract}

%
\vspace{2pc}

\section{Introduction}

\label{sec: Introduction}
\normalsize

Innovation is an idea, method, technology, or process, that is perceived as novel~\cite{oguntuase2020academic}. It is an evolutionary process that typically relates to improvements over existing ideas or solutions. The importance of studying the generative mechanism of innovation lies in understanding the infrastructure necessary to improve the system's pe rformance and efficiency~\cite{hildrum2014turning}. Innovation becomes increasingly difficult if no new elements are added to the toolbox of scientists, artists, entrepreneurs, and other innovators, since it requires a combination of expertise, skills, and ideas to create novel knowledge and products. In recent years, interdisciplinarity has been encouraged across organisations to leverage the potential benefits of complementary expertise, aiming to solve increasingly complex social and economic problems~\cite{arthur2021foundations,arthur2009nature,anderson2018economy,chuluun2017firm, begovic2021power}. In this context, social interactions and collaboration between individuals or organisations (i.e.~agents) generating complex social networks are fundamental to provide means to exchange and combine information from diverse sources. The non-linear benefits of social interactions are known to have an impact on innovation~\cite{Bettencourt2007, rocha2021scaling}.

Models of innovation have mostly focused on studying the adoption of innovative ideas or products by groups of agents~\cite{montanari2010spread,moolenaar2010social, kuandykov2010impact,ting2009simulate}. Diffusion of innovation is a social process governed by the impact of media and social interactions~\cite{karsai2014complex}. The relationship between agents, the number of fragmented or disconnected clusters, population density, and population heterogeneity are among environmental factors affecting information exchange~\cite{barrat2008dynamical, centola2007complex}. Modeling innovation in networks makes it possible to understand the impact of social interactions diffusion. Recent studies have shown how ideas spread and how diffusion depends on the network structure~\cite{guardiola2002modeling, martins2009opinion, iacopini2018network}. A simple model was proposed to study how the trade-off between acquiring new skills and improving existing skills shapes social networks~\cite{smolla2019cultural}, while the relation between network structure and product and process innovation was analysed using configurational terms~\cite{ozkan2016configuration}. The impact of the network structure in respect to population size and connectivity was studied by simulating innovation and diffusion of cultural traits in populations with stereotyped social structures~\cite{cantor2021social}. These studies show that small, seemingly insignificant idiosyncrasies of their structures can heavily impact innovation's diffusion among members of a social network~\cite{abrahamson1997social,muller2019effect}.

The creation of innovation is different from the spread of innovation. The creation of innovation concerns the evolution of agent and group knowledge or the evolutionary process of increasing the complexity of ideas, products, or services by building up simpler existing solutions~\cite{derex2018divide, smolla2019cultural}. Previous studies on social sciences hypothesise that social contacts within groups constrain the information flow and create knowledge holes (structural holes) across the network~\cite{burt1992structural, burt2004structural}. Individuals or organisations connecting different social groups are believed to have higher potential of innovation because they can leverage knowledge from different groups. Empirical studies have provided evidence on the correlation between social network structure and creativity~\cite{ozkan2016configuration,derex2018divide, fang2010balancing, derex2016partial}. Correlation studies are however unable to explain the mechanisms connecting the network structure to the potential of innovation creation, and the interplay of social network and innovation dynamics at various structural scales.

In this paper, we build upon the hypothesis that network structure regulates innovation dynamics via social interactions, and study the impact of social network structures, on the innovation creation process. We assume that more complex technologies (i.e.~innovation) result from simpler technologies, either via self-creation when sufficient knowledge is available to an agent, or via social learning when there is exchange of knowledge between agents. We devise a mechanistic model incorporating the creation of technologies and social structure where information exchange only occurs between connected agents with a sufficiently similar knowledge base. We study the effect of microscopic (local or agent level) and macroscopic (global or group level) network structures of both theoretical and empirical networks on the creation of new technologies and the ability of agents and groups of agents to create technologies with higher complexity. We find a paradoxical effect of social network structure. While agents benefit of local network heterogeneity, this heterogeneity affects negatively the overall innovation ability of the group. Furthermore, reducing the openness to interdisciplinary collaboration removes the benefit of network heterogeneity on innovation.

\section{Materials and Methods} \label{sec: Model}

\subsection{Innovation model}

\begin{figure*}[ht!]
    \centering
    \includegraphics[scale= 0.5]{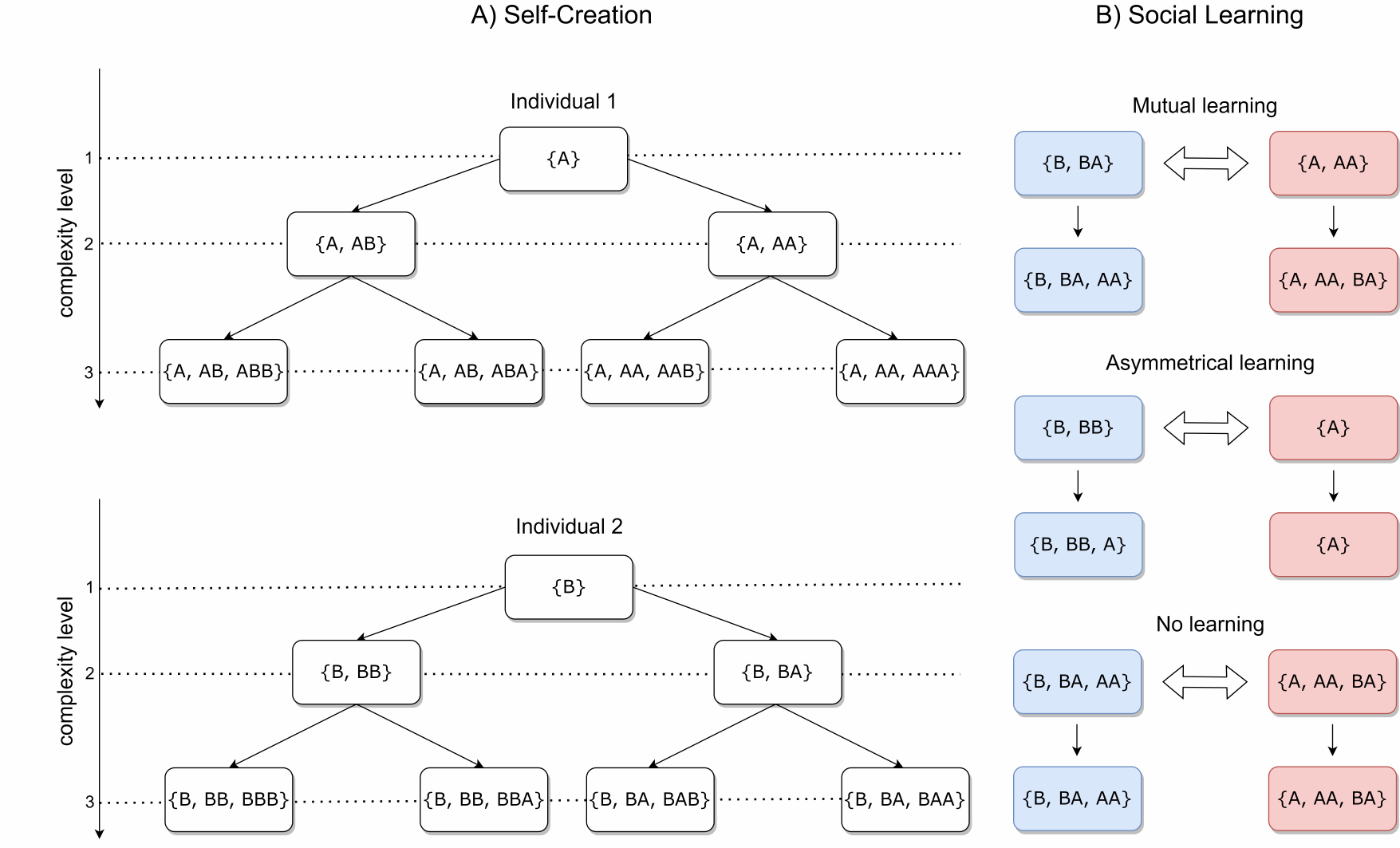}
     \centering\caption{Innovation process. (A) Self-creation: A new and more complex technology can be created from a simpler technology (e.g.\ $C=2$ from $C=1$). (B) Social learning: (i) mutual learning, both agents learn the most complex technology of each other; (ii) asymmetrical learning, one agent -blue here- has a complexity level larger than the red agent and thus only the blue agent learns; (iii) no learning, both agents have the same level of complexity but there is an overlap of the most complex technologies, hence no learning for both agents.}
    \label{fig:01}
\end{figure*}

In our model, we assume a population of $N$ agents where agents $i$ and $j$ are connected via a social tie $(i,j)$. Each agent $i$ has a set $S_i(t)$ of technologies at time $t$. A technology is defined as a combination of letters $A$ and $B$ such that the complexity level $C^h$ of a technology $h$ is given by the number of letters representing the respective technology. The complexity of an agent $C_i(t)$ is defined by the technology with the highest complexity in the set $S_i(t)$. The size of $S_i(t)$ ($=D_i(t)$) represents the diversity of technologies known by agent $i$. New technologies are created via a branching process (Fig.~\ref{fig:01}A). The simplest technologies are either $A$ or $B$ ($C=1$). An agent can create one out of two new technologies $AA$ or $AB$ ($C=2$) from $A$ (or $BB$ or $BA$ from $B$). The same process is valid for more complex technologies ($C>2$). This mechanism creates a historical dependency of innovation with various possible trajectories for $S_i(t)$~\cite{derex2018divide}. Another way of increasing the size of $S_i(t)$ is via social learning (Fig.~\ref{fig:01}B). In our model, an agent can learn only the most complex technology of its neighbour, provided that its complexity is not higher than the agent's complexity level. An agent $i$ can learn a technology with complexity level $C^h$ only if $C^h \leq C_i(t)$. In the first scenario of social learning, both agents have the same level of complexity and can learn from each other (mutual learning). In the second scenario, the most complex technology of the blue agent is $BB$ $(C_{\text{blue}}=2)$ but the red agent has complexity $C_{\text{red}} = 1$. Therefore learning only happens from agent red to blue (asymmetrical learning). In the third scenario, the most complex technologies of the two agents are similar and thus they do not learn from each other (no learning). During social learning, because an agent $i$ cannot learn technologies with higher complexity, only its diversity ($D_i(t)$) may increase. Self-creation, on the other hand, may lead to increase in both the complexity level and diversity of agents.

Initially, all agents have $S_i(t=0) = \{ A \}$ or $S_i(t=0) = \{ B \}$ technologies distributed uniformly at random (i.e.~$D_i(t=0) = |S_i(t=0)| = 1$). At each time step $t$, an agent can increase $S_i(t)$ through 2 mechanisms: (i) create one new technology based on its own knowledge of simpler technologies (self-creation) or (ii) acquire one technology via social learning. We select an agent uniformly at random to create a new technology with probability $p_i$. This probability is adjusted to impose that agents must know a sufficient number of accumulated knowledge to be able to create one that is more complex than those in its own portfolio. The sigmoid function shows the mathematical equivalent of this probability.
\begin{equation}
    \label{Sigmoid}
    p_i(t) = \frac{1}{1+e^{-\beta (D_i(t)-\alpha C_i(t))}}
\end{equation}
In our simulation we have set $\alpha=1.2$. Increasing $\alpha$ makes the network to quickly stabilize and not grow much in complexity level. Reducing $\alpha$ leads to a stable state where the complexity increases to relatively high number for long times. We set $\beta= 30$ to obtain a step function, such that there is a large probability of creating a new technology if $D_i(t)>\alpha C_i(t)$, whereas this probability is small if $D_i(t)<\alpha C_i(t)$.

If there is no self-innovation (i.e.~with probability $1-p_i(t)$), the agent $i$ interacts with one of its social contacts $j$ (chosen at random) to collaborate and share knowledge. For the exchange between agents $i$ and $j$ to be successful, two conditions have to be met: (i) the technological similarity $\theta_{ij}(t)$,
\begin{equation}
    \label{Similarity}
    \theta_{ij}(t) = \dfrac{| S_i(t) \cap S_j(t) |}{|S_i(t) \cup S_j(t)|}
\end{equation}
should be above a threshold $\theta_h$ (which is a hyperparameter of the model); and (ii) an agent $i$ can only learn a technology from $j$ that is equally or less complex than its own level of complexity, i.e.\ $C_j(t) \le C_i(t)$. Therefore, the complexity level of an agent does not increase through social learning, whereas its portfolio of technologies, i.e.\ its diversity, does. Higher diversity increase the agent's chance of self-innovation the next time it is selected. This process is repeated until the system reaches the stationary, defined as the time when  $\frac{d\langle C \rangle}{dt}<10^{-6}  $.

\subsection{Social Networks}

To study the effect of social network structure on the creation of technologies, we use theoretical network models and real-world network data. A network is defined by a set of $N$ nodes in which nodes $i$ and $j$ are connected by a link $(i,j)$. In our model, a node $i$ corresponds to an agent $i$ (an individual or an organisation). The number of social contacts of $i$ is the degree $\kappa_i$. The clustering coefficient $cc_i= 2e_i/(\kappa_i (\kappa_i-1))$ (where $e_i$ is the number of links between common neighbours of $i$) gives the fraction of triangles connected to node $i$. The betweenness centrality $b_i$ gives the fraction of all shortest-paths, between any two nodes $j$ and $k$ in the network ($\sigma_{jk}$), passing through node $i$, i.e.~$b_i= \sum _{i \ne j \ne k} \frac{\sigma_{jk} (i)}{ \sigma_{jk}}$.

The reference theoretical network model is the Erd\"os-R\'enyi random model (ER) in which links are made between pairs of nodes with a fixed probability $q$. The emerging structure is homogeneous with a characteristic node degree~\cite{erdHos1960evolution}. The second network model contains heterogeneity in the clustering coefficient of nodes, to capture different levels of clustering in the network. We use a configuration model to generate random networks where the distribution of clustering coefficients is controlled and the rest of the structure is randomised for a given average degree~\cite{miller2009percolation, newman2009random}. The third theoretical network uses the configuration model with a fixed degree distribution $p(\kappa) \propto \kappa^{-\beta}$, with $\beta=2.5$ for finite mean and variance~\cite{newman2003structure}.
 In our simulations, we first generate $10$ realisations of each network model with the same set of parameters, and run $10$ times the innovation model to obtain averages over $m=100$ points.
 
We also apply our model of creation of technologies to three data sets of social collaboration. The first data set is a co-authorship network of scientists working on network (COA). This network is a one-mode projection from the bipartite graph of authors and their scientific publications. The network is formed by nodes  representing authors connected by links $(i,j)$ if there is at least one joint publication~\cite{newman2006finding}.
The second empirical data set corresponds to current and previous affiliation of researchers at 206 computer science departments at universities in the USA (COM). Each link represents that a professor working at university $i$ got a PhD at university $j$, therefore, this network represents the flow of human capital between universities ~\cite{clauset2015systematic}.
The third data set comes from Scopus and represents scientific collaborations between institutions in New Zealand in the period from 2010 to 2015 (SCOP). The network is formed by nodes representing institutions (e.g.\ universities, organisations), connected by links if there is at least one joint publication with authors from both institutions~\cite{aref2018analysing}.

Finally, to study the correlations in the network structure, we randomise the links of SCOP to create random networks with the same average degree but fixed clustering (CSCOP) and fixed degree (DSCOP) distributions. Table~\Ref{Table1} shows a summary of all networks used in our study.

\begin{table}[htbp]
\centering
\small 
\begin{tabular}{
   |p{0.15\linewidth}
  |p{0.55\linewidth}
  |p{0.1\linewidth}
  |p{0.05\linewidth}|
  }
 \hline
Code & \centering \raggedright Network model & $N$ & $\langle \kappa \rangle$ \\
\hhline{|=|=|=|=|}
 ER  & \centering \raggedright Erd\"os-R\'enyi random model & 1000 & 10  \\
\hline
 RC & \centering \raggedright Random clustered model & 1000 & 10  \\
\hline
 SF & \centering \raggedright Scale-free random model & 1000 & 10  \\
\hline
COA & \centering \raggedright Collaborations between authors & 1461 & 4  \\
\hline
COM & \centering \raggedright Current and previous affiliation of researchers & 206 & 27  \\
\hline
SCOP & \centering \raggedright Collaborations between institutions & 1511 & 6 \\
\hline
CSCOP& \centering \raggedright Randomised version of SCOP with fixed Clustering &  1511 & 6 \\
\hline
DSCOP& \centering \raggedright Randomised version of SCOP with fixed degree & 1511 & 6 \\
\hline
\end{tabular}
\caption{Summary of the main characteristics of the network models used in our study.}
\label{Table1}
\end{table}

\section{Results}
\label{sec:Results}

In this section, we will study the impact of social network structure on innovation at the macroscopic (network level) and microscopic (agent level) scales. We first analyse the evolution of innovations on random network models to study specific structures and then apply the model on real-world collaboration networks to understand the effect of real structures on the innovation dynamics. Finally, we analyse correlations in real networks by studying randomised versions of real networks with a chosen fixed structure.

\subsection{Group innovation}

\begin{figure*}[!ht]
    \centering
    \includegraphics[scale=0.5]{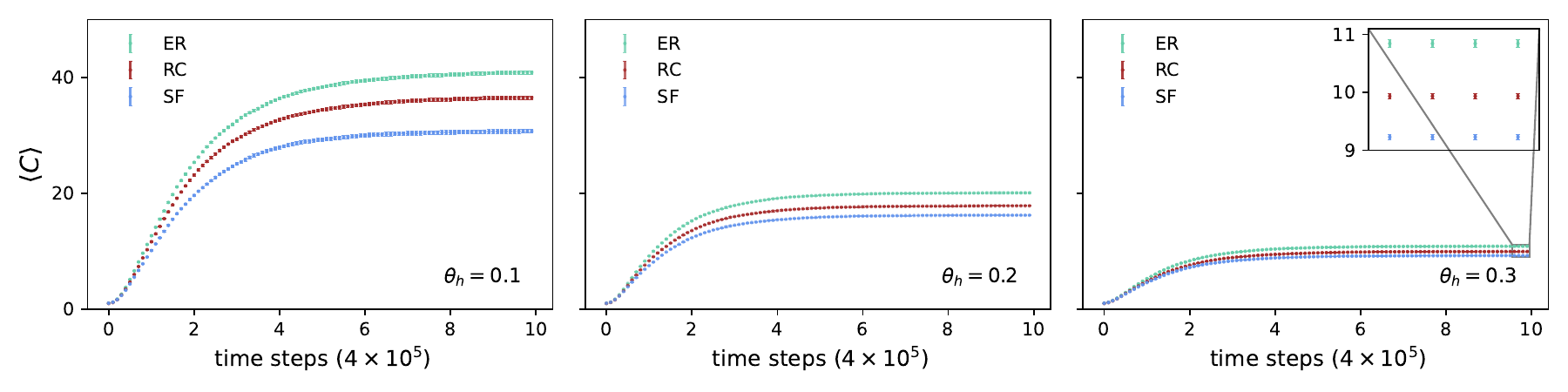}
    \includegraphics[scale=0.5]{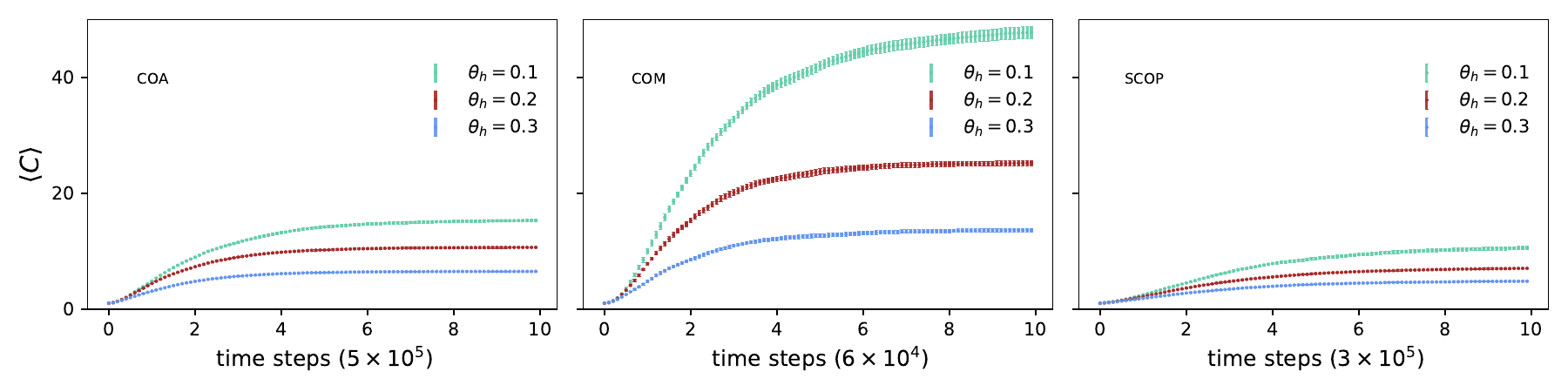}
    \centering\caption{Temporal evolution of the group complexity $\langle C \rangle$ for different thresholds of similarity $\theta_h$ and networks (see Table~\Ref{Table1}). The averages are taken over $m=100$ realisations of the network and random starting conditions. Bars represent standard error.}
    \label{figTotC}
\end{figure*}

\begin{figure}
    \centering \includegraphics[scale= 0.5]{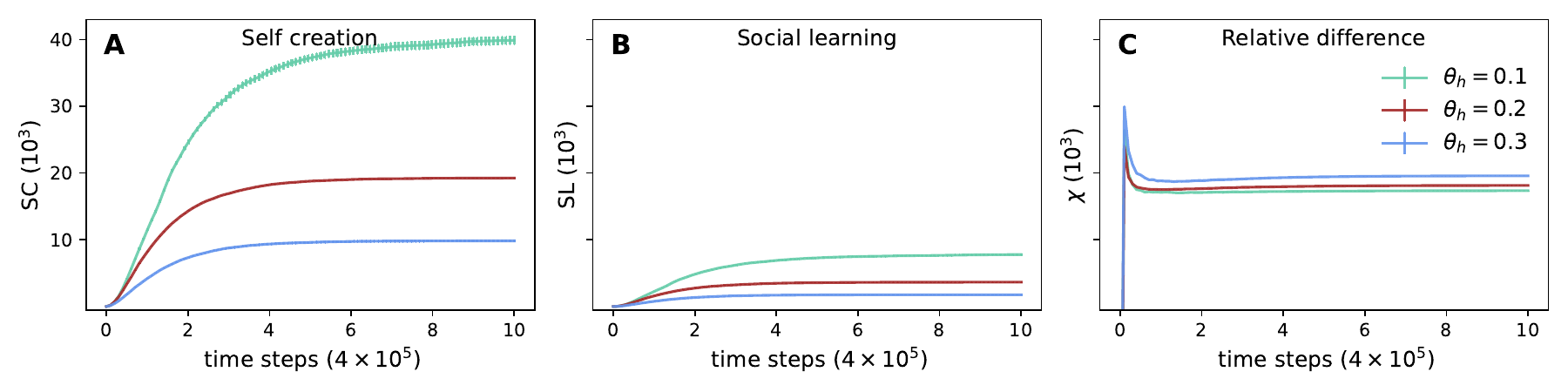}
        \centering \caption{ Temporal evolution of the number of times that agents in the network (A) self-create  $ \left (SC\right )$ and (B) social learn $\left ( SL\right )$ a new technology; (C) The relative importance of $SC$ to $SL$, $ \left (\chi= 100~\frac{SC-SL}{SC+SL}\right )$, for the ER network model. The averages are taken over $m=100$ realisations of the network and random starting conditions.
    \label{fig:counter}}
\end{figure}

\begin{figure*}
    \centering
    \includegraphics[scale= 0.8]{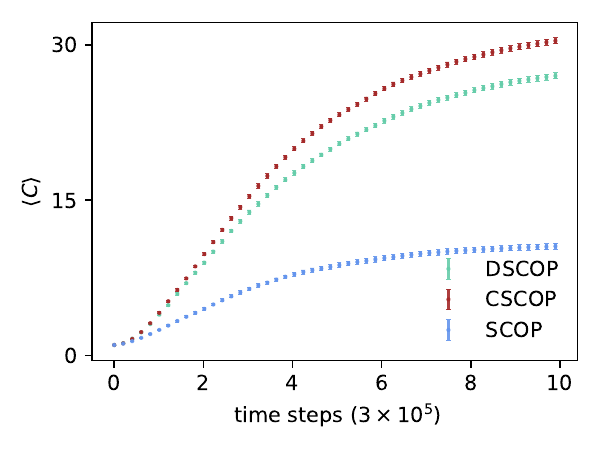}
    \centering
    \caption{Temporal evolution of the group complexity $\langle C \rangle$ on empirical structures (Table~\ref{Table1}). The averages are taken over $m=10$ realisations of random starting conditions.}
    \label{fig:realComplexity_time}
\end{figure*}

Figure~\ref{figTotC}(A-C) show the temporal evolution of the average complexity $\langle C \rangle$ for all $N$ agents for three network models: Erd\"os-R\'enyi (ER), random clustered (RC), and scale-free (SF). In all cases, a fast increase of $\langle C \rangle$ in the beginning is followed by a slower increase until the system reaches a nearly-stationary state. This happens because in the early stages, creating new complex technologies is relatively easier since agents have a small set of simple technologies and diversity grows faster than complexity ($D_i(t)>C_i(t)$), providing the conditions for self-creation (See Methods and Eq.~\Ref{Sigmoid}). Furthermore, there is higher overlap of knowledge of contacts (because $D_i(t)$ is small during early times) and thus exchanges via social learning are facilitated. As the dynamics evolves, the difference between diversity and complexity decreases and innovation becomes less likely. The diversity of the set of technologies also increases and less overlap of knowledge is observed between social contacts, reducing the speed of innovation, until it eventually eventually converges.

Figure~\Ref{fig:counter} shows the temporal evolution of the cumulative number of times per agents that they increase their number of technologies via self-creation ($SC$), and social learning ($SL$). Also and the relative difference between $SC$ and $SL$ ($\chi=100~\frac{SC-SL}{SC+SL}$), for a random network structure (ER model) with various similarity thresholds $\theta_h$, (results are qualitatively similar for other network configurations). In the early stages, agents can social learn and self-create new complex technologies. After a while, the difference in the complexity level of agents increases, and agents with higher complexity cannot teach but only learn new technologies from their social contacts (Fig.~\ref{fig:01}B), leading to a decrease in social learning. Once these agents learn all the complex technologies of their social contacts, only self-creation is possible. At the later stage, diversity is not sufficiently large for self-creation and the system reaches stationarity. The results also show that for a larger similarity threshold, social learning is less important to transfer new technologies but important to increase the diversity of the agents.

The average complexity $\langle C \rangle$ is higher for homogeneous networks (ER, the reference case) in comparison to heterogeneous structures. The heterogeneous clustering reduces  $\langle C \rangle$ (RC) however the degree heterogeneity (SF) has a higher effect on limiting group innovation (Fig.~\ref{figTotC}A-C). When the network has structural heterogeneity, the input information is not the same for all agents. Therefore the agents achieve different levels of complexity, and this disparity decreases social learning and consequently  $\langle C \rangle$.
Figure~\ref{figTotC}(D-F) shows the temporal evolution of $\langle C \rangle$ for the empirical networks, that are highly heterogeneous in terms of clustering, degree, and betweenness centrality. In all cases, the complexity increases relatively fast in the beginning, and eventually reaches stationarity. Social learning increases with an increase in the threshold $\theta_h$, also leading to an increase in innovation and thus on $\langle C \rangle$ .

The results also show the importance of the similarity between social contacts to promote group innovation. A lower level of similarity in known technologies between contacts ($\theta = 0.1$, i.e.~a higher level of interdisciplinarity between social contacts) generates on average more complex technologies (Fig.~\ref{figTotC}a) in comparison to a higher level of similarity ($\theta_h=0.3$, i.e.~less interdisciplinarity)(Fig.~\ref{figTotC}c). This result means that innovation is less likely if social learning occurs between too similar agents. On one hand, a high level of similarity is necessary for two agents to communicate effectively and exchange knowledge, on the other hand, too similar agents are unable to fill gaps and complement each other's knowledge, limiting their ability to create new complex technologies. Our model indicates that reducing the threshold $\theta_h$ (i.e.~promote interdisciplinarity) is an effective means to avoid saturation of knowledge and generate new technologies. For higher $\theta_h$, the social network structure becomes less relevant because the chance of having similar agents decreases and the connectivity does not affect social learning.

Figure~\ref{fig:realComplexity_time} compares the evolution of innovation using one empirical network (SCOP) and the randomised versions (CSCOP and DSCOP, see Table~\ref{Table1}) to show the impact of correlated structures in the dynamics. The result shows that, besides heterogeneity, the correlation between edges in a network also reduces group innovation.

\subsection{Microscopic dynamics}

The group dynamics is a result of microscopic interactions between the agents. Social networks have different levels of at the individual (i.e.~local) level heterogeneity that can be captured by various network measures. We study 3 local structures: (i) the degree $\kappa_i$; (ii) the betweenness centrality $b_i$, and (iii) the clustering coefficient $cc_i$ of agent $i$ (See Methods). 

\begin{figure*}[ht!]
    \centering
    \includegraphics[scale= 0.5]{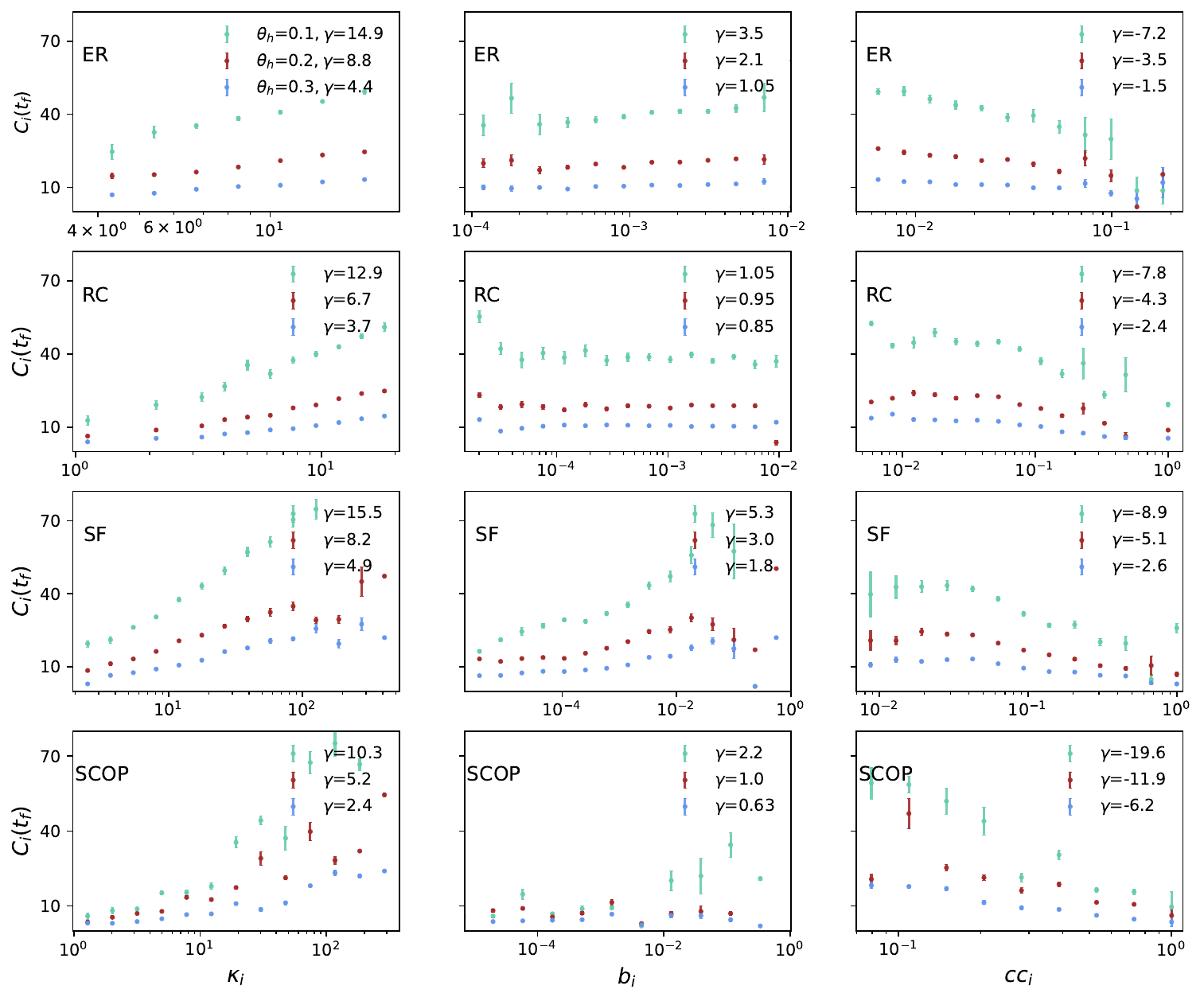}
    \centering
    \caption{Individual complexity and network structures. Agent complexity $C_i(t_f)$ and local network structures for various network models (Table~\Ref{Table1}) at stationarity. Panels (A,D,G,J) correspond to degree $\kappa_i$; (B,E,H,K) to betweenness centrality $b_i$; and (C,F,I,L) to clustering coefficient $cc_i$. The coefficient $\gamma$ indicates the slope of the relation between $C_i(t_f)$ and $\log(\kappa_i)$, $\log(b_i)$ and $\log(cc_i)$. The averages are taken over $m=100$ realisations of the network and random starting conditions. Bars represent standard error. 
}
    \label{figCom}
   
\end{figure*}

Figure~\ref{figCom} shows the connection between the complexity level of individual agents and their network features. Independently of the network, we observe that agent the complexity increases with its degree (Fig.~\ref{figCom}A,D,G,J) and betweenness centrality (Fig.~\ref{figCom}B,E,H,K) whereas it decreases with its clustering (Fig.~\ref{figCom}C,F,I,L). The effect of the degree is stronger in the SF model because of the degree heterogeneity. Similarly, the RC model has stronger clustering heterogeneity than the other models and thus the effect on the agent complexity level is higher. Empirical networks have structural correlations but also in this case, one observes similar trend between local connectivity and complexity level (Fig.~\ref{figCom}J,K,L). The degree indicates the local level of connectivity of an agent and thus its potential to be influenced by multiple social contacts, or network neighbours. A higher social degree thus increases the probability of getting new technologies because of higher exposure. The betweenness centrality measures the brokerage potential of an agent connecting socially diverse groups of network nodes, and therefore, its potential to leverage knowledge from multiple groups of agents. The clustering coefficient is a local measure of embeddedness in a social group. Agents that are overly embedded in a social group are less exposed to diverse views and knowledge because clustering tends to homogeneise the information circulating within the group.

The similarity threshold $\theta_h$ affects social learning. If $\theta_h$ is high, agents must be highly similar for interactions to result in social learning of new complex technologies. Figure~\ref{figCom} shows that the lack of openness or willingness to inter-disciplinary collaboration (i.e.~high $\theta_h$) not only reduces the agent level of complexity but also the relevance of social heterogeneity in the dynamics of social learning. Openness to collaborate with contacts with different knowledge (low $\theta_h$), on the other hand, leverages the diversity provided by social heterogeneity to increase innovation. This can be seen in the values of $\gamma$ which is the regression coefficient of the complexity level $C_i(t_f)$ to the logarithm of the network measure $\log(x_i \in \left\{\kappa_i, b_i, cc_i\right\} ) $. The importance of openness is highlighted by the fact that very open agents with few contacts may achieve the same level of complexity as less open agents with many contacts. Table \Ref{Table2} shows the range of $\gamma$ for three random network structure (ER, RC, and SF) which have the same number of nodes and average degree. By decreasing $\theta_h$, the effect of network structure on innovation increases. The coefficient $\gamma$ is not fixed and depends on the network structure at both local and global scales. However, it is always inversely related to $\theta_h$.

\begin{table}[htbp]
\centering
\begin{tabular}{
  |p{\dimexpr.12\linewidth-2\tabcolsep-1.3333\arrayrulewidth}
  |p{\dimexpr.18\linewidth-2\tabcolsep-1.3333\arrayrulewidth}
   |p{\dimexpr.18\linewidth-2\tabcolsep-1.3333\arrayrulewidth}
  |p{\dimexpr.18\linewidth-2\tabcolsep-1.3333\arrayrulewidth}|
  }
 \hline
\diagbox{$\theta_h$ }{$x$}  &  ~~~~~$\kappa_i$ &  ~~~~~$b_i$ & ~~~~~$cc_i$\\
\hline
 0.1 & [12.9 , 15.5] & [1.05 , 5.3] & [-8.9 , -7.2] \\
\hline
0.2 & [6.7 , 8.8] & [0.95 , 3.0] & [-5.1 , -3.5] \\
\hline
0.3 & [3.7 , 4.9] & [0.85 , 1.8] & [-2.6 , -1.5] \\
\hline
\end{tabular}
\caption{
\label{Table2}
Dependency between the network measures $x_i \in \{ \kappa_i, b_i, cc_i \}$ and the complexity levels $C_i(t_f)$ of the individual agents in the stationary regime as fitted by the function $C_i(t_f) = C_0 + \gamma \log (x_i)$. The range of the fitted $\gamma$ values is given for the three random networks (ER, RC, and SF).   
 }
\end{table}

\begin{figure*}[ht!]
    \centering
       \includegraphics[scale= 0.37]{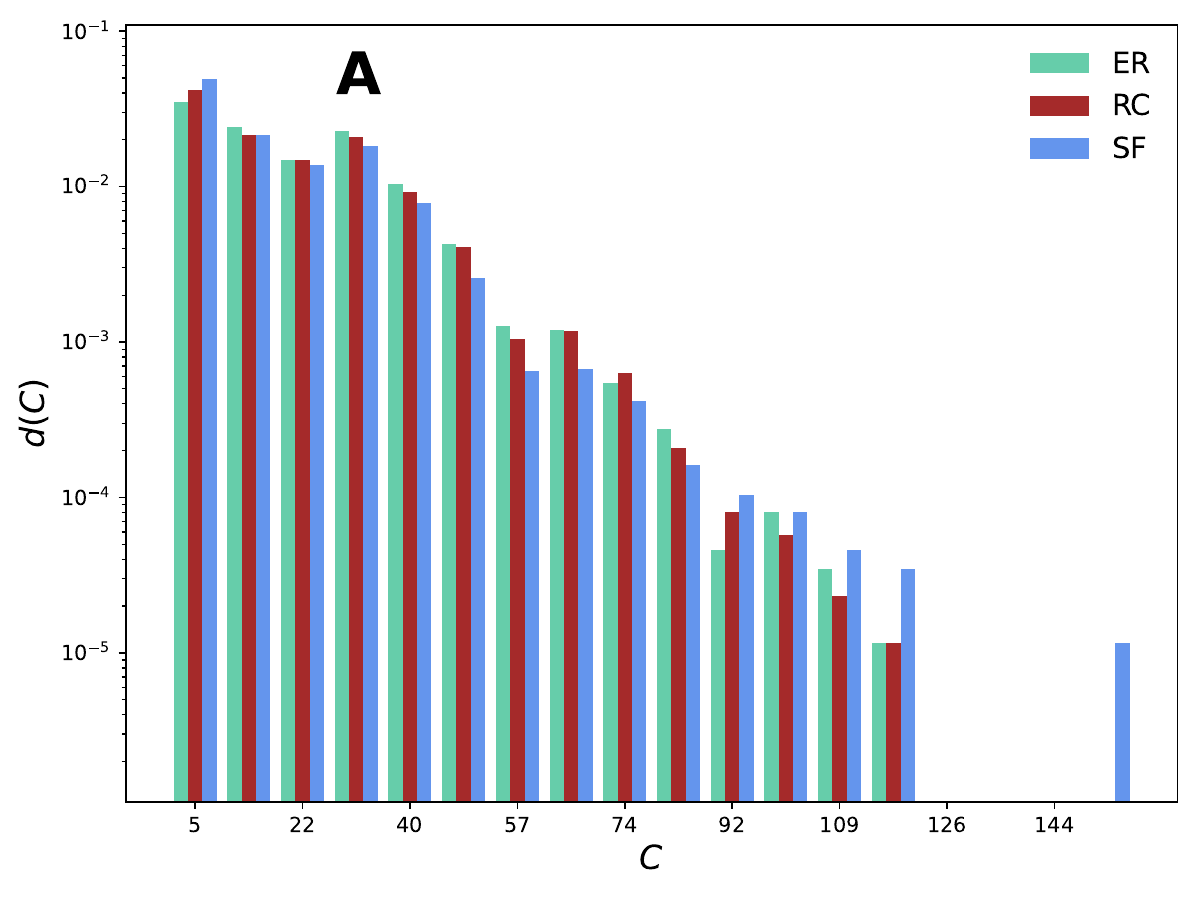}
       \includegraphics[scale= 0.37]{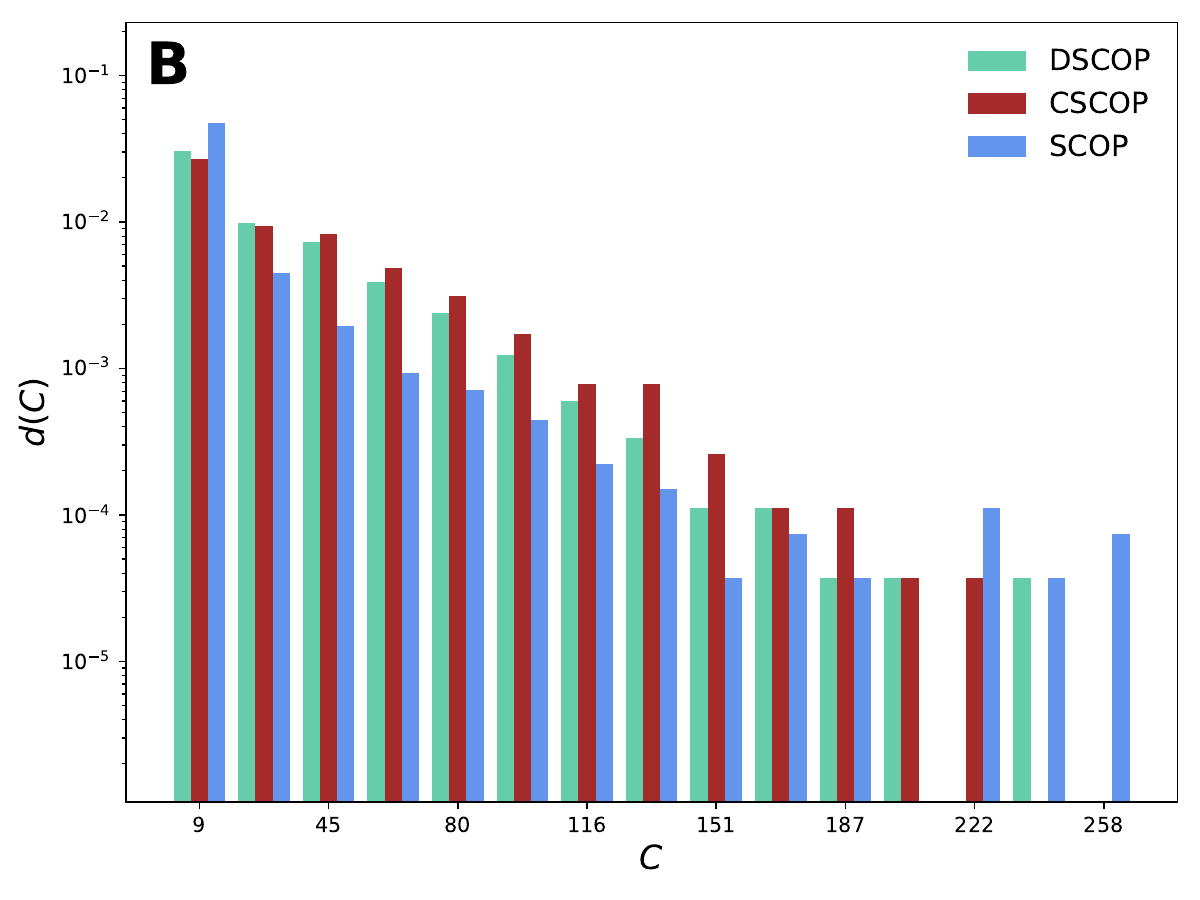}
       
       \centering\caption{Distribution of individual complexity level. ($d(C)$) at stationarity ($t_f$) for A) random networks ($t_f= 4 \times 10^5$), and B) the empirical network (SCOP) and its randomised versions (CSCOP and DSCOP), ($t_f=3 \times 10^5$). We used $\theta_h= 0.1$ in these simulations. The averages are taken over $m=50$ starting conditions.}
    \label{fig:distribution}
    
\end{figure*}

Network heterogeneity not only distributes the ability of agents to be innovative but also affects top performers. Figure~\ref{fig:distribution} shows the distribution density of the agent complexity level in the stationary state. The fraction of agents with low complexity is lower in homogeneous networks (ER) in comparison to heterogeneous networks (RC and SF). As the complexity increases, only small differences are observed for different networks. However, the highest levels of individual complexity are achieved with the degree (SF) and clustered (RC) heterogeneous models. The SF structure promotes the emergence of some top performers and relatively many lower performers (i.e.~higher inequality) followed by the clustered networks. In contrast, a random network structure promotes more equal social opportunities and innovation potential (i.e.~less inequality). Figure~(\Ref{fig:distribution}B) shows the distribution density of complexity at stationarity for the SCOP, CSCOP and DSCOP networks. The result indicates that homogeneity causes top performers to disappear. In the network with the highest average complexity (DSOP) the number of top performers is lower than in other networks and the distribution of the individual agent complexity is more homogeneous.

\subsection{Knowledge normalisation}

The diversity of technologies $D_i(t)$ of each agent has to be larger than the complexity $C_i(t)$ (Eq.~\Ref{Sigmoid}) to create a new technology. Social learning only happens when a sufficient variety of knowledge exists between social contacts, otherwise, existing knowledge is reinforced and new technologies are not created. Social learning is a dynamic process that promotes innovation but also normalises (i.e.~homogenises) local knowledge. This normalisation can be measured by the similarity $\Omega_i(t)$ in the set of technologies between an agent and its social contacts.

\begin{equation}
    \label{eqSim}
    \Omega_i(t)= \frac{1}{\kappa_i} \sum_{j=1}^{\kappa_i} \theta_{ij}(t)
\end{equation}

\begin{figure*}[ht!]
    \centering
    \includegraphics[scale= 0.5]{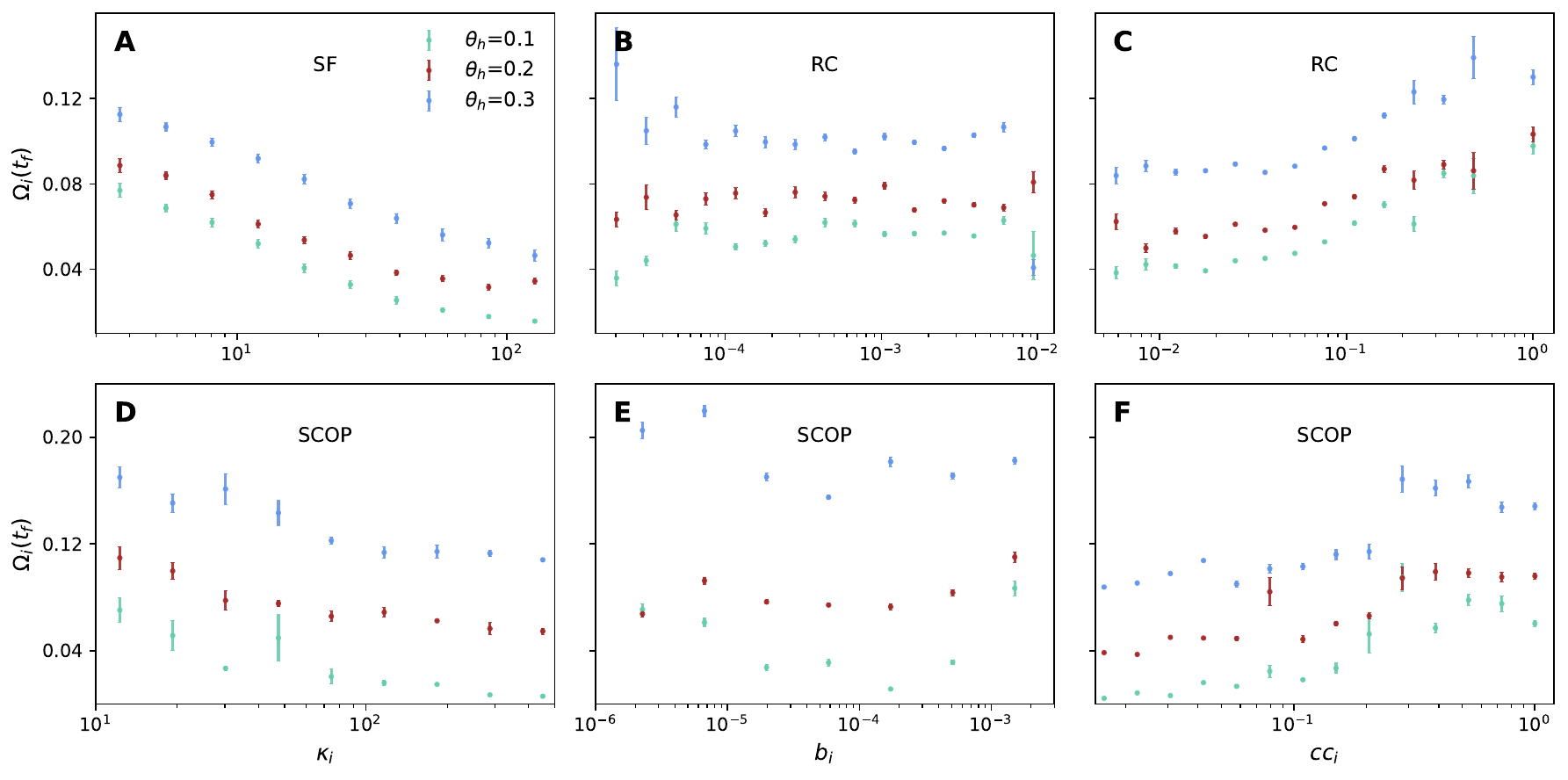}

    \caption{Knowledge similarity and network structure. The similarity $\Omega_i(t_f)$ of an agent $i$ and local network structure, (A) $\kappa_i$ (SF network); (B) $b_i$ (RC network); (C) $cc_i$ (RC network); (D) $\kappa_i$ (SCOP network); (E) $b_i$ (SCOP network); (F) $cc_i$ (SCOP network). Bars represent standard error.}
    \label{figSim}.
\end{figure*}

Figure~\ref{figSim} shows $ \Omega_i(t_f)$ in different networks in comparison to local network measures. There is anti-correlation between the level of complexity of each agent and the similarity with their social contacts. Over time, agents become more similar which eventually saturates the possibilities to generate more complex technologies. Although too similar agents do not exchange  new technologies, an agent that is too different from its social contacts ($\theta < \theta_h$) cannot benefit from social learning either. An agent that innovates more becomes less similar to its social contacts. Therefore, the similarity shows an inverse relationship with complexity, meaning that an agent with higher degree or higher betweenness centrality has a lower level of similarity to its social contacts. On the other hand, high clustering makes social learning inside the cluster more frequent than between clusters. Increasing $\theta_h$ leads to higher similarity between social contacts because only agents with high similarity can learn from each other whereas the openness given by low $\theta_h$ promotes diversity.

\section{Conclusion}
\label{sec:Conclusion}

Innovation is a result of combining different ideas, skills, or knowledge to create new technologies, ideas, or solutions. To increase the innovation potential, individuals and organisations may interact and exchange information via social learning. The structure of the social networks thus affects the rate of innovation. In this paper, we devised and tested a mechanistic model, that includes an evolutionary process of innovation whereby agents interact via social networks and create more complex technologies from simpler ones. Within the model we studied the interplay between network structure and the creation of technologies at the individual and group scales.

We found that local structural heterogeneity increases the ability of individual agents to generate more complex technologies. There is a strong correlation between degree and betweenness centrality and the level of agent complexity. Being embedded in social clusters reduces the ability of an agent to innovate. This happens because more central agents collaborate with various agents or social groups, and thus are influenced by diverse sources of different knowledge whereas clusters tend to homogenise knowledge within the group, thus reducing opportunities to leverage the advantages of diversity. We found however a contrasting effect at different scales. While such inequality benefits individual agents, the same structural heterogeneity reduces the innovation potential of the group. This happens because heterogeneity implies that only a few agents are sufficiently exposed to diverse knowledge in detriment of the majority that are exposed to similar knowledge, either because of less connectivity or because it is not bridging clusters of knowledge. Once central agents develop high levels of complexity, those connected to them become less similar to the central agents and thus have lower chances to innovate via social learning.

Heterogeneity is maybe not good for the group but still, top-performers in the heterogeneous networks are better than the average innovation in homogeneous cases. In other words, heterogeneity in the network structure on the one hand causes heterogeneity in the innovation of agents, which reduces their average innovation relative to the homogeneous network, but on the other hand causes some agents to have much higher performance than the average of the homogeneous group.

If there is no willingness or ability to learn from interdisciplinary collaborations (i.e.~from agents with different sets of knowledge), the overall level of complexity of the group is negatively affected and decreases. Furthermore, the heterogeneity of the social structure is only beneficial at the microscopic level when there is sufficient openness to interdisciplinarity. If agents avoid interdisciplinarity, the requested level of similarity between contacts is too high and the social network structure becomes irrelevant.

Although network heterogeneity may be advantageous for individual agents, this is not true for the group. According to our model, promoting more collaboration of less connected agents or reducing the centrality of highly central agents increases not only the overall performance of the social network but also the overall diversity of technologies generated by the collaborative network. This model has a limitation on the mechanism of social learning, due to the assumption that high complexity agents are not able to teach social contacts who are at a lower complexity level, rendering them less useful for fostering group innovation. One consequence is that innovation may increase if organisations and governments provide more support to reduce heterogeneity, for example by supporting less central agents to improve their position in the network or by reducing “the-rich-get-richer” schemes~\cite{bol2018matthew} that benefit well-connected agents (i.e.~by making research funding less dependent on candidate track record, signaled by past funding or past publications, or a more random distribution of resources).

Our model does not account for agent abilities and resources to exploit social learning. We assume that all agents process information and are able to create technologies at the same speed. High performers may achieve higher levels of innovation with fewer contacts and vice-versa. By adding a temporal dimension to our model, we might be able to include such mechanisms to better describe the agent and social processes necessary to adapt the network structure to maximise innovation~\cite{gross2008adaptive, Aoki2016}. Our model for innovation does not also account for shocks and paradigm shift. Agents could suddenly create a technology that has a very higher level in comparison to their knowledge.  By adding a jump term to our model for self-creation, we might be able to include such sudden innovation and adapt our model to the real contexts.

\section*{Acknowledgments}
LECR thanks Matthew Smith for pointing out helpful references.

\section*{References\label{except}}
\bibliographystyle{ieeetr}
\bibliography{References}

\begin{thebibliography}{10}

\bibitem{oguntuase2020academic}
O.~J. Oguntuase, ``Academic entrepreneurship, bioeconomy, and sustainable development,'' in {\em Handbook of Research on Approaches to Alternative Entrepreneurship Opportunities}, pp.~32--57, IGI Global, 2020.

\bibitem{hildrum2014turning}
J.~M. Hildrum, ``Turning stone into gold and silver into stone: On the importance of studying innovation,'' {\em The Innovation Journal}, vol.~19, no.~2, p.~1, 2014.

\bibitem{arthur2021foundations}
W.~B. Arthur, ``Foundations of complexity economics,'' {\em Nature Reviews Physics}, vol.~3, no.~2, pp.~136--145, 2021.

\bibitem{arthur2009nature}
W.~B. Arthur, {\em The nature of technology: What it is and how it evolves}.
\newblock Simon and Schuster, 2009.

\bibitem{anderson2018economy}
P.~W. Anderson, {\em The economy as an evolving complex system}.
\newblock CRC Press, 2018.

\bibitem{chuluun2017firm}
T.~Chuluun, A.~Prevost, and A.~Upadhyay, ``Firm network structure and innovation,'' {\em Journal of Corporate Finance}, vol.~44, pp.~193--214, 2017.

\bibitem{begovic2021power}
B.~Begovi{\'c}, ``The power of creative destruction: Economic upheaval and the wealth of nations by philippe aghion, c{\'e}line antonin, and simon bunel,'' {\em Panoeconomicus}, vol.~68, no.~4, pp.~577--586, 2021.

\bibitem{Bettencourt2007}
L.~M.~A. Bettencourt, J.~Lobo, D.~Helbing, C.~K\"uhnert, and G.~West, ``Growth, innovation, scaling, and the pace of life in cities,'' {\em Proceedings of the National Academy of Sciences USA}, vol.~104, no.~17, pp.~7301--7306, 2007.

\bibitem{rocha2021scaling}
L.~E.~C. Rocha, J.~Ryckebusch, K.~Schoors, and M.~Smith, ``The scaling of social interactions across animal species,'' {\em Scientific Reports}, vol.~11, no.~1, pp.~1--10, 2021.

\bibitem{montanari2010spread}
A.~Montanari and A.~Saberi, ``The spread of innovations in social networks,'' {\em Proceedings of the National Academy of Sciences}, vol.~107, no.~47, pp.~20196--20201, 2010.

\bibitem{moolenaar2010social}
N.~M. Moolenaar and P.~J. Sleegers, ``Social networks, trust, and innovation. how social relationships support trust and innovative climates in dutch schools,'' {\em Social Network Theory and Educational Change}, pp.~97--114, 2010.

\bibitem{kuandykov2010impact}
L.~Kuandykov and M.~Sokolov, ``Impact of social neighborhood on diffusion of innovation s-curve,'' {\em Decision Support Systems}, vol.~48, no.~4, pp.~531--535, 2010.

\bibitem{ting2009simulate}
Z.~Ting, G.~Baojun, and X.~Huiyu, ``Simulate the effects of advertising on the diffusion of innovation with cellular automata,'' {\em Science and Technology Progress and Policy}, vol.~1, 2009.

\bibitem{karsai2014complex}
M.~Karsai, G.~Iniguez, K.~Kaski, and J.~Kert{\'e}sz, ``Complex contagion process in spreading of online innovation,'' {\em Journal of The Royal Society Interface}, vol.~11, no.~101, p.~20140694, 2014.

\bibitem{barrat2008dynamical}
A.~Barrat, M.~Barthelemy, and A.~Vespignani, {\em Dynamical processes on complex networks}.
\newblock Cambridge University Press, 2008.

\bibitem{centola2007complex}
D.~Centola and M.~Macy, ``Complex contagions and the weakness of long ties,'' {\em American Journal of Sociology}, vol.~113, no.~3, pp.~702--734, 2007.

\bibitem{guardiola2002modeling}
X.~Guardiola, A.~Diaz-Guilera, C.~J. Perez, A.~Arenas, and M.~Llas, ``Modeling diffusion of innovations in a social network,'' {\em Physical Review E}, vol.~66, no.~2, p.~026121, 2002.

\bibitem{martins2009opinion}
A.~C.~R. Martins, C.~d.~B. Pereira, and R.~Vicente, ``An opinion dynamics model for the diffusion of innovations,'' {\em Physica A: Statistical Mechanics and its Applications}, vol.~388, no.~15-16, pp.~3225--3232, 2009.

\bibitem{iacopini2018network}
I.~Iacopini, S.~Milojevi{\'c}, and V.~Latora, ``Network dynamics of innovation processes,'' {\em Physical Review Letters}, vol.~120, no.~4, p.~048301, 2018.

\bibitem{smolla2019cultural}
M.~Smolla and E.~Ak{\c{c}}ay, ``Cultural selection shapes network structure,'' {\em Science advances}, vol.~5, no.~8, p.~eaaw0609, 2019.

\bibitem{ozkan2016configuration}
E.~Ozkan-Canbolat and A.~Beraha, ``Configuration and innovation related network topology,'' {\em Journal of Innovation and Knowledge}, vol.~1, no.~2, pp.~91--98, 2016.

\bibitem{cantor2021social}
M.~Cantor, M.~Chimento, S.~Q. Smeele, P.~He, D.~Papageorgiou, L.~M. Aplin, and D.~R. Farine, ``Social network architecture and the tempo of cumulative cultural evolution,'' {\em Proceedings of the Royal Society B}, vol.~288, no.~1946, p.~20203107, 2021.

\bibitem{abrahamson1997social}
E.~Abrahamson and L.~Rosenkopf, ``Social network effects on the extent of innovation diffusion: A computer simulation,'' {\em Organization Science}, vol.~8, no.~3, pp.~289--309, 1997.

\bibitem{muller2019effect}
E.~Muller and R.~Peres, ``The effect of social networks structure on innovation performance: A review and directions for research,'' {\em International Journal of Research in Marketing}, vol.~36, no.~1, pp.~3--19, 2019.

\bibitem{derex2018divide}
M.~Derex, C.~Perreault, and R.~Boyd, ``Divide and conquer: intermediate levels of population fragmentation maximize cultural accumulation,'' {\em Philosophical Transactions of the Royal Society B: Biological Sciences}, vol.~373, no.~1743, p.~20170062, 2018.

\bibitem{burt1992structural}
R.~S. Burt, {\em Structural holes}.
\newblock Harvard University Press, 1992.

\bibitem{burt2004structural}
R.~S. Burt, ``Structural holes and good ideas,'' {\em American journal of sociology}, vol.~110, no.~2, pp.~349--399, 2004.

\bibitem{fang2010balancing}
C.~Fang, J.~Lee, and M.~A. Schilling, ``Balancing exploration and exploitation through structural design: The isolation of subgroups and organizational learning,'' {\em Organization Science}, vol.~21, no.~3, pp.~625--642, 2010.

\bibitem{derex2016partial}
M.~Derex and R.~Boyd, ``Partial connectivity increases cultural accumulation within groups,'' {\em Proceedings of the National Academy of Sciences}, vol.~113, no.~11, pp.~2982--2987, 2016.

\bibitem{erdHos1960evolution}
P.~Erd{\H{o}}s and A.~R{\'e}nyi, ``On the evolution of random graphs,'' {\em Publications of the Mathematical Institute of the Hungarian Academy of Sciences}, vol.~5, no.~1, pp.~17--60, 1960.

\bibitem{miller2009percolation}
J.~C. Miller, ``Percolation and epidemics in random clustered networks,'' {\em Physical Review E}, vol.~80, no.~2, p.~020901, 2009.

\bibitem{newman2009random}
M.~E.~J. Newman, ``Random graphs with clustering,'' {\em Physical Review Letters}, vol.~103, no.~5, p.~058701, 2009.

\bibitem{newman2003structure}
M.~E.~J. Newman, ``The structure and function of complex networks,'' {\em SIAM Review}, vol.~45, no.~2, pp.~167--256, 2003.

\bibitem{newman2006finding}
M.~E.~J. Newman, ``Finding community structure in networks using the eigenvectors of matrices,'' {\em Physical Review E}, vol.~74, no.~3, p.~036104, 2006.

\bibitem{clauset2015systematic}
A.~Clauset, S.~Arbesman, and D.~B. Larremore, ``Systematic inequality and hierarchy in faculty hiring networks,'' {\em Science advances}, vol.~1, no.~1, p.~e1400005, 2015.

\bibitem{aref2018analysing}
S.~Aref, D.~Friggens, and S.~Hendy, ``Analysing scientific collaborations of new zealand institutions using scopus bibliometric data,'' in {\em Proceedings of the Australasian Computer Science Week Multiconference}, pp.~1--10, 2018.

\bibitem{bol2018matthew}
T.~Bol, M.~de~Vaan, and A.~van~de Rijt, ``The matthew effect in science funding,'' {\em Proceedings of the National Academy of Sciences}, vol.~115, no.~19, pp.~4887--4890, 2018.

\bibitem{gross2008adaptive}
T.~Gross and B.~Blasius, ``Adaptive coevolutionary networks: a review,'' {\em Journal of the Royal Society Interface}, vol.~5, no.~20, pp.~259--271, 2008.

\bibitem{Aoki2016}
T.~Aoki, L.~E.~C. Rocha, and T.~Gross, ``Temporal and structural heterogeneities emerging in adaptive temporal networks,'' {\em Physical Review E}, vol.~93, p.~040301, 2016.

\end{thebibliography}

\end{document}